\documentclass[aip, rsi, amsmath, amssymb, preprint]{revtex4-1}

\usepackage{graphicx}
\usepackage{color}
\usepackage{dcolumn}
\usepackage{bm}

\begin{document}


\title{From the difference of structures to the structure of the difference}

\author{Massimiliano Zanin}
 \email{massimiliano.zanin@ctb.upm.es.}
\affiliation{Centro de Tecnolog\'ia Biom\'edica, Universidad Polit\'ecnica de Madrid, 28223 Madrid, Spain}
\affiliation{Departamento de Engenharia Electrot\'ecnica, Faculdade de Ci\^ecias e Tecnologia, Universidade Nova de Lisboa, 2829-516 Caparica, Portugal}

\author{Ernestina Menasalvas}
\affiliation{Centro de Tecnolog\'ia Biom\'edica, Universidad Polit\'ecnica de Madrid, 28223 Madrid, Spain}

\author{Xiaoqian Sun}
\author{Sebastian Wandelt}
\affiliation{School of Electronic and Information Engineering, Beihang University, 100191 Beijing, China}
\affiliation{Beijing Key Laboratory for Network-based Cooperative ATM, 100191 Beijing, China}

\date{\today}

\begin{abstract}
When dealing with evolving or multi-dimensional complex systems, network theory provides with elegant ways of describing their constituting components, through respectively time-varying and multi-layer complex networks. Nevertheless, the analysis of how these components are related is still an open problem.
We here propose a framework for analysing the evolution of a (complex) system, by describing the structure created by the difference between multiple networks by means of the Information Content metric. As opposed to other approaches, as for instance the use of global overlap or entropies, the proposed one allows to understand if the observed changes are due to random noise, or to structural (targeted) modifications. 
We validate the framework by means of sets of synthetic networks, as well as networks representing real technological, social and biological evolving systems.
We further propose a way of reconstructing network correlograms, which allow to convert the system's evolution to the frequency domain.
\end{abstract}



\maketitle

\section{Introduction}
\label{sec:Intro}

Although complex networks theory \cite{strogatz2001exploring, newman2003structure} was initially used to describe the structure underpinning individual complex systems, in recent years there has been an explosion in the number of situations in which (potentially large) sets of networks have to be studied in a comparative way.
The availability of multiple related networks may be the natural result of analysing different, yet compatible systems - as, for instance, functional brain networks obtained from a large set of healthy people, with the aim of identifying common connectivity patterns \cite{greicius2003functional}; or from control subjects and patients suffering from a given condition \cite{seeley2009neurodegenerative}, to detect differences between them.
This can nevertheless also stem from the analysis of a single system across its parameters' and temporal dimensions. Following on the previous example, neuroscientists may be interested in characterising the temporal evolution of such networks during a long cognitive task \cite{bassett2011dynamic, zalesky2014time}, or across different frequency bands \cite{buldu2017frequency}.
Potential examples are not limited to neuroscience, and indeed appear in all research fields in which complex networks have been applied \cite{costa2011analyzing}, {\it i.e.} across social, biological and technological systems - a clear example of the latter being air transport networks~\cite{neal2014devil, belkoura2016multi}.

The analysis of the differences between two or more networks is a two-fold problem. On one hand, it entails the quantification of such differences \cite{brodka2017quantifying}, {\it e.g.} by calculating a set of topological metrics and by comparing their normalised values \cite{costa2007characterization}; and, on the other hand, the understanding of the dynamical processes causing such changes.
These two aspects of the problem are orthogonal, as both of them have to be taken into account for the correct understanding of an observed evolution. The fact that two networks are not equal does not imply the presence of a structured evolutionary process, as they may be the result of describing the same system under observational noise. Such conclusion cannot be drawn even from a statistically significant change in some topological metric: {\it e.g.} a reduction in the modularity may be the result of a random link rewiring, but also of a targeted process aimed at disrupting the modular structure. Even an increase in modularity may be the result of a random process, albeit with low probability. Lastly, and on the same line, one should not correlate the magnitude of the changes with the presence of targeted processes: noise does not necessarily result in small fluctuations only.
These two aspects, {\it i.e.} description and structureness, are also of high relevance of real-world applications. For instance, in the specific case of brain functional networks, the presence of an unstructured difference between control subjects and patients may be ascribed to a global loss of brain connectivity, while structured changes may suggest a focused reorganisation of the information flow. 

The second previously discussed point, {\it i.e.} the understanding of the dynamical processes causing a change, is a specific aspect of the more general problem known as {\it phenotype to genotype} \cite{strohman2002maneuvering, zanin2016phenotype}. While we can observe only the phenotype of a system, in this case the resulting physical or functional network, what we would really like to understand is the genotype that has created it. If several phenotypes are available, {\it e.g.} we can observe the temporal evolution of the system, we can in principle use the phenotype's dynamics to (partly) reconstruct the genotype: in other words, we can use the ``difference of structures'' to unveil the underlying ``structure creating such difference''.

Inspired by this, we here present a framework designed to answer the following specific question: {\it do the observed changes follow a structure, or are they simply the result of random fluctuations?}
This framework is based on {\it a}) the calculation of the {\it difference} between the two observed networks, {\it b}) the representation of such difference as a new {\it difference network}, and {\it c}) the analysis of its structural characteristics. Specifically, we start from the assumption that changes resulting from non-random processes are characterised by correlations, which reflect in the presence of a meso-scale in the difference network. Such meso-scale can then be detected using a broad-band topological metric, {\it i.e.} the {\it Information Content} \cite{zanin2014information}, and its significance assessed through a statistical test based on ensembles of equivalent random networks. By means of a set of synthetic evolving networks, we show that this approach outperforms other alternatives, as the ones based on cross-network correlations \cite{bianconi2013statistical} or von Neumann entropy \cite{braunstein2006laplacian, braunstein2006some}.
We further demonstrate the usefulness of the proposed solution by analysing three real systems, respectively technical (the evolution of the world-wide air transport network), social (human contact networks in a hospital) and biological (comparison of functional brain networks corresponding to different frequency bands).
We conclude this work by showing how this approach can be used to construct a {\it network correlogram}, which, among others, can be used to detect the natural frequency of a time-evolving network.


\section{Metric definition}
\label{sec:metric}

\subsection{Information Content}
\label{sec:IC}

For the sake of completeness, we here include a short overview of the {\it Information Content} metric, which is the basis of the proposed methodology. For a more complete description, the reader may refer to Ref. \cite{zanin2014information}.

The rationale behind the definition of the {\it Information Content} is that a regular network, or more generally any network presenting a meso-scale structure, displays strong correlations between the node's connectivity patterns. The information encoded by pairs of such correlated nodes is thus redundant, as the connections of one of them almost completely define the second one's. A clear example is yielded by networks with a strong community structure, in which two nodes belonging to the same community usually share most of their neighbours.

Following this idea, and given an initial network, the proposed algorithm identifies the pair of nodes whose merging would suppose the smallest information loss, {\it i.e.} that share most of their connections. The analysis of two nodes $i$ and $j$ thus entails, firstly, the creation of a vector of differences $m$, with $m_k = 1 - \delta_{a_{i, k}, a_{j, k}}$ and $\delta$ being the Kronecker Delta. Secondly, the information encoded by $m$ is assessed through the classical Shannon's entropy, defined as:

\begin{equation}
	I_{i, j} = -2N( p_0 \log_2 p_0 + p_1 \log_2 p_1 ),
\end{equation}

$p_0$ and $p_1$ being respectively the frequency of zeros and ones in $m$, and $N$ the number of nodes in the network. Note that $I_{i, j}$ represents the quantity of information required to reconstruct $j$'s connections given $i$'s ones; and thus the quantity of information lost when both nodes are merged.

The pair of nodes minimising $I$ are then merged, and the quantity of information lost in the process is approximated by $I$. The process is iteratively repeated, until one single node remains, being the final Information Content $IC$ the sum of the information lost in all steps. As shown in a previous work \cite{zanin2014information}, low IC values indicate the presence of some kind of regularity in the link arrangement, including communities, hubs, or core-periphery configurations.

\subsection{Comparing two networks}
\label{sec:Comparing}

Suppose two networks, each one described by a corresponding adjacency matrix $\mathcal{A}_1$ and $\mathcal{A}_2$, which have been observed under different conditions. Firstly, the most simple case includes two independent networks, representing two different systems - albeit of the same size, {\it i.e.} the same number of nodes. Secondly, these adjacency matrices can represent different layers of a multiplex network \cite{boccaletti2014structure}. Finally, the networks may represent different snapshots of the same time-evolving system \cite{holme2012temporal}. In all cases, changes between $\mathcal{A}_1$ and $\mathcal{A}_2$ can be encoded in a matrix $\mathcal{D} = |\mathcal{A}_1 - \mathcal{A}_2|$, whose element $d_{i, j}$ is equal to $1$ when the corresponding link has changed in the two analysed networks, and zero otherwise. Note that $\mathcal{D}$ can be interpreted as the adjacency matrix of a network whose links depict a corresponding change between $\mathcal{A}_1$ and $\mathcal{A}_2$.

With respect to the meso-scale structure of the difference network $\mathcal{D}$, only two situations can be encountered. First, changes between $\mathcal{A}_1$ and $\mathcal{A}_2$ can be random, for instance due to measurement noise, or more generally due to uncorrelated forces; $\mathcal{D}$ would then resemble the adjacency matrix of a random network. Second, if changes between $\mathcal{A}_1$ and $\mathcal{A}_2$ are somehow correlated, the resulting network should present some kind of meso-scale structure. For instance, if changes only affect the connections of one node, $\mathcal{D}$ will be star-like shaped. All intermediate situations, {\it e.g.} with only a part of the links modified at random, can be interpreted as a special (and noisy) case of the latter situation.

If changes are not random, and thus are correlated and form a meso-scale structure, the latter should be detected by the $IC$ metric. An algorithm for the comparison of different networks can thus be designed, composed of the following steps: {\it i}) calculate $\mathcal{D}$ as $|\mathcal{A}_1 - \mathcal{A}_2|$; {\it ii}) calculate the $IC$ of the network $\mathcal{D}$; {\it iii}) compare $IC(\mathcal{D})$ with the value obtained in an ensemble of equivalent random networks.
As for the latter point, several ways of normalising the obtained value are available. Firstly, one can simply calculate:

\begin{equation}
IC^* = \frac{ IC(\mathcal{D}) }{ \mu(IC_r) },
\end{equation}

where $\mu(IC_r)$ is the average {\it Information Content} obtained in an ensemble of random networks, with the same number of nodes and links as $\mathcal{D}$. Note that $IC^*$ usually takes values in $(0, 1)$, with values close to one indicating a random structure of the network $\mathcal{D}$, and thus a random difference between $\mathcal{A}_1$ and $\mathcal{A}_2$; and $0 < IC^* < 1$, the presence of a structure in the changes. Further note that, while $IC^* > 1$ is possible, it would indicate a structure more random than a random network, and can thus only be the result of random fluctuations.

While $IC^*$ provides a quantitative assessment of the structure of changes, it yields little information about the statistical significance of the same. In order to tackle this issue, a normalisation based on a Z-Score can be used:

\begin{equation}
IC^\dagger = \frac{ IC(\mathcal{D}) - \mu(IC_r) }{ \sigma(IC_r) }.
\end{equation}

$IC^\dagger$ values close to zero indicate random modifications between $\mathcal{A}_1$ and $\mathcal{A}_2$, while negative values indicate modifications driven by some structure. The advantage of this formulation is that $IC^\dagger$ can easily be transformed into a $p$-value, provided $IC_r$ follows a normal distribution - condition that is not fulfilled only for very small random networks.

\begin{figure*}[!tb]
\begin{center}
\includegraphics[width=0.9\textwidth]{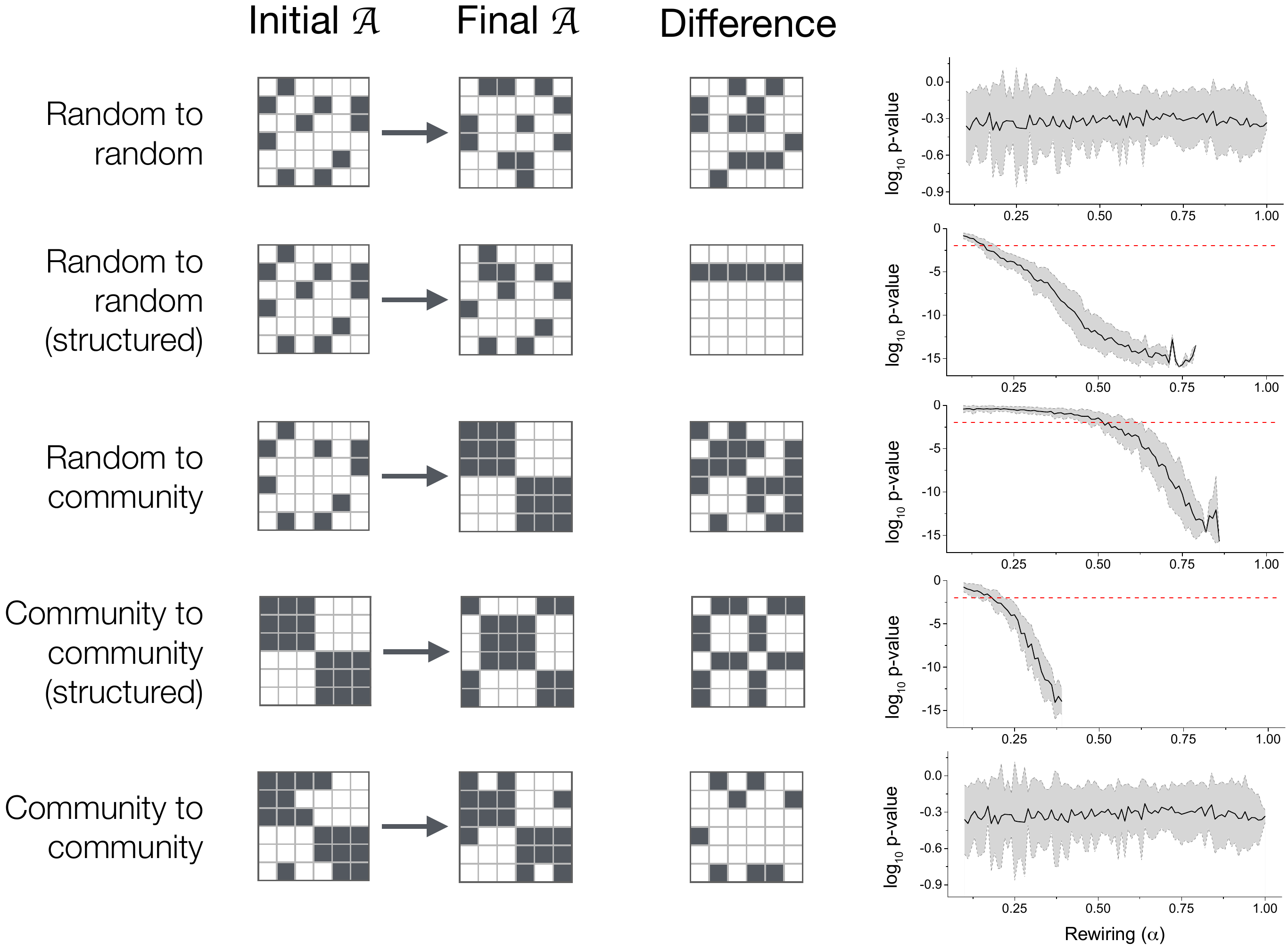}
\caption{Examples of the application of the proposed algorithm to synthetic evolving networks. From left to right, the four columns respectively represent: the initial adjacency matrix; the final adjacency matrix; the difference matrix; and the evolution of the $\log_{10}$ of the $p$-value of the test assessing the presence of a structured change, as a function of the rewiring $\alpha$ - see main text for details. All calculations have been executed with $100$-nodes networks, while the depicted adjacency matrices have a smaller size for the sake of clarity. The black line and grey bands on the right hand side respectively depict the average and $1 \sigma$ band, as obtained in $100$ random realisations.}\label{fig:01}
\end{center}
\end{figure*}

\subsection{Validation on synthetic networks}

A simple way of validating the proposed algorithm involves the use of a set of controlled evolutions, {\it i.e.} governed by rules ensuring that the start and end points are known topologies. Given these two networks $\mathcal{A}_{start}$ and $\mathcal{A}_{end}$, we construct a third network $\mathcal{A}$ whose links are drawn from $\mathcal{A}_{end}$ with probability $\alpha$, and from $\mathcal{A}_{start}$ with probability $1 - \alpha$; and finally compare $\mathcal{A}$ with the initial network $\mathcal{A}_{start}$.
Note that, for $\alpha = 0$, $\mathcal{A} = \mathcal{A}_{start}$ and $\mathcal{D} = 0_{N \times N}$; on the other hand, $\alpha = 1$ implies that $\mathcal{A} = \mathcal{A}_{end}$ and $\mathcal{D} = |\mathcal{A}_{end} - \mathcal{A}_{start}|$. Therefore, $\alpha$ controls the degree of morphing between $\mathcal{A}_{start}$ and $\mathcal{A}_{end}$.

Several evolutions of interest are analysed in Fig. \ref{fig:01}. The four columns, from left to right, respectively represent the initial (rewiring $\alpha = 0$) and final ($\alpha = 1$) networks; $\mathcal{D}$, for the maximum rewiring $\alpha = 1$; and the evolution of the $\log_{10}$ of the $p$-value of $IC^\dagger$, as a function of the rewiring $\alpha$, calculated between the original and the rewired network. While, for the sake of clarity, the depicted adjacency matrices have a small size, all results have been obtained with networks of $100$ nodes and $100$ random realisations.

The first row describes the rewiring of a random network into a second random one. As there is no correlation nor structure between the links that have changed, the resulting matrix $\mathcal{D}$ presents a random connectivity and no meso-scale; consequently, the drop in $IC$ never becomes statistically significant, as depicted in the right panel.
The second example, while similar, presents an important difference: if both the initial and final networks are random, the second is obtained by reversing the set of neighbours of one single node - see the corresponding matrix $\mathcal{D}$. Note that, in this case, while the initial and final points are random, the evolution process is a structured one. This is correctly detected by the proposed metric, with the $p$-value dropping below $0.01$ for $\alpha \approx 0.15$.

Similar behaviours are observed in the third and fourth examples, which describe two different networks converging towards a community structure. As creating or modifying a community requires links to be activated and de-activated in a targeted way, the metric detects the presence of a meso-scale in $\mathcal{D}$.
Finally, the latter example consists of a situation in which both the starting and final networks have the same community structure, being both contaminated by random noise. Accordingly, the difference between both has a random nature, and the $p$-value never becomes statistically significant.

Some general conclusions can be drawn from these results. Firstly, and most importantly, the structure of the two networks $\mathcal{A}_1$ and $\mathcal{A}_2$ is not relevant; instead, only the changes that are required to evolve from the former to the latter are. Specifically, two completely random networks may be associated with a structured change between them; and two well-structured networks may differ in a random fashion.
Secondly, the presence of a statistically significant structured process is the result of the trade-off between the fraction of modified links and their organisation. For instance, it is worth noting that in the second example of Fig. \ref{fig:01} a statistical significant result is reached for $\alpha \approx 0.15$, as all modified links belong to the same node, while an $\alpha > 0.5$ is required in the third example. In other words, the change of a few strongly correlated links can be as significant as the change of many links, when the relationship between them is weaker.

\subsection{Comparison with other approaches}
\label{sec:OtherM}

Among the literature dealing with the problem of complex network comparison \cite{brodka2017quantifying}, two alternative approaches are worth considering: the comparison of network topological properties on one hand, and of the raw adjacency matrices on the other.

The former approach is the most common: one or more metrics, synthesising the topology of the networks, are calculated and compared. The main advantage is that it allows to compare heterogeneous (in terms of number of nodes) networks, provided the metrics are normalised against equivalent random networks. As explained in the introduction, this is not equivalent to what here proposed, as such comparison cannot explain the nature of the evolutionary process.
 
This section thus focuses in the latter option, {\it i.e.} on strategies for directly comparing two or more adjacency matrices, specifically through the use of correlations and entropy measures. We show how there are several situations in which those strategies' effectiveness is below what yielded by the proposed approach.

\subsubsection*{Correlation}

An interesting, and yet simple way of comparing two networks, or two layers in a multiplex network, is to calculate the correlation between the links present in both of them. In other words, given two networks $\mathcal{A}_1$ and $\mathcal{A}_2$, the correlation expresses the probability that if $a^{\mathcal{A}_1}_{i, j} = 1$, then $a^{\mathcal{A}_2}_{i, j} = 1$. More generally, one can calculate a global overlap $O^{\mathcal{A}_1, \mathcal{A}_2}$ as the total number of pair of nodes connected at the same time by a link in networks $\mathcal{A}_1$ and $\mathcal{A}_2$, as proposed in Ref.~\cite{bianconi2013statistical}, {\it i.e.}:

\begin{equation}
O^{\mathcal{A}_1, \mathcal{A}_2} = \sum _{i < j} a ^{\mathcal{A}_1}_{i, j} a ^{\mathcal{A}_2}_{i, j}.
\label{eq:O}
\end{equation}

Eq. \ref{eq:O} can further be normalised by considering the number of links present in both networks.
It has recently been shown \cite{cellai2013percolation, baxter2016correlated} that such global overlap has important implications in the percolation, and thus in the robustness, of multiplex networks - as the presence of correlated (redundant) links slows down the disruption of the giant component of the network under random link removal.

Two extreme situations can be encountered when considering the correlation between two networks: when all links are equal in both networks, and thus the correlation is maximal; and when links are reciprocal, {\it i.e.} $a ^{\mathcal{A}_1}_{i, j} = 1 - a^{\mathcal{A}_2}_{i, j}$, yielding a maximally negative correlation. $\mathcal{D}$ would respectively be a null and a complete matrix, and $IC^\dagger = 0$ in both cases - in other words, a strong structure drives the evolution between both networks.
More interesting situations arise in the middle range, {\it i.e.} when only part of the links are different.
To illustrate, let us consider the situation depicted in the first two rows of Fig. \ref{fig:01}, and suppose the initial and final matrices $\mathcal{A}$ are random and have the same link density of $0.5$. In both cases, $O^{\mathcal{A}_1, \mathcal{A}_2} \approx 0.25$ (as half of the activated links are expected to coincide); yet, the two resulting $IC$ values are completely different. This proves that the global overlap metric $O$ does not provide information on the underlying mechanism driving such difference, as a same correlation value may be the result of random or structured changes.

\subsubsection*{von Neumann entropy}

The von Neumann entropy ($S_{VN}$) is a metric that was initially introduced in quantum mechanics to assess the degree of mixing of the quantum states encoded in a probability distribution - and hence in a density matrix $\rho$. While the concept of a state probability distribution is not defined for complex networks, the metric can still be calculated over any density matrix, {\it i.e.} any Hermitian and positive semidefinite matrix. As previously shown \cite{braunstein2006laplacian, braunstein2006some}, $S_{VN}$ can be calculated over the density Laplacian matrix as:

\begin{equation}
S_{VN} = - \mathtt{Tr} \frac{\mathcal{L}}{<k>N} \log \frac{\mathcal{L}}{<k>N},
\end{equation}

where $k$ is the average degree, $N$ the number of nodes composing the network, and $\mathcal{L} = \mathcal{D} - \mathcal{A}$ the corresponding Laplacian matrix. The von Neumann entropy has been demonstrated to be a good quantifier of the regularity of a network structure, with higher values obtained in graphs with uniform degree distributions, and smaller values in heterogeneous networks \cite{passerini2009quantifying}. In a way similar to our approach, $S_{VN}$ has been used to compare different networks \cite{de2016spectral}, but with the limitations discussed below.

Let us suppose two networks with the same number of nodes and links, $\mathcal{A}_r$ and $\mathcal{A}_m$, respectively having a random and a modular structure. More specifically, the elements of the former adjacency matrix are drawn from a binomial distribution $\mathcal{B}(1, 0.5)$, while those of the second are defined as $a_{i, j} = 1$ for $i < N/2, j < N/2$ and $i > N/2, j > N/2$ (and zero otherwise).
In the limit of large values of $N$, both networks are characterised by the same expected value of the degree distribution, {\it i.e.} $0.5$, and all nodes are further expected to have the same (or very similar) degree $N / 2$. Due to the dependency of $S_{VN}$ on the degree distribution, both networks are expected to have similar values of the entropy.

\begin{figure*}[!tb]
\begin{center}
\includegraphics[width=0.8\textwidth]{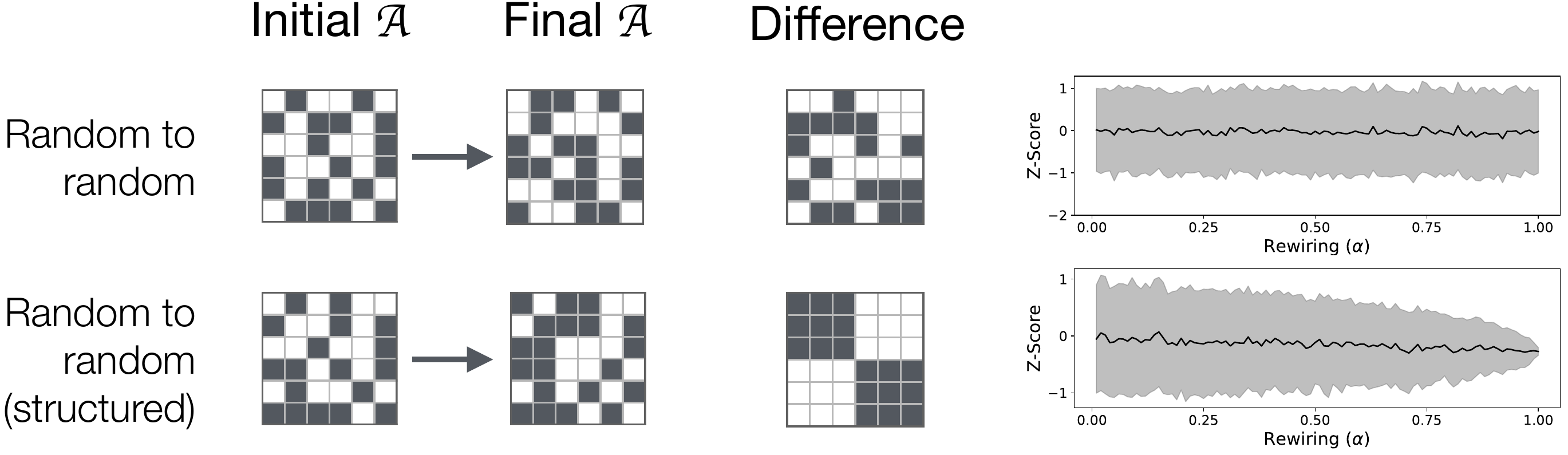}
\caption{Evolution of the Z-Score of the von Neumann entropy, as a function of the rewiring $\alpha$, and for two synthetic networks. Panel meanings are as in Fig. \ref{fig:01}.}\label{fig:01a}
\end{center}
\end{figure*}

It is easy to construct situations in which the difference network $\mathcal{D}$ is equal to $\mathcal{A}_r$ or $\mathcal{A}_m$. For instance, starting from a random network with a link density of $0.5$, the first case is obtained when this is compared with another random network with the same size and link density; on the other hand, the second case is obtained by inverting the activation of links in the upper left and bottom right quarters of the adjacency matrix. The behaviour of the von Neumann entropy in these two situations is depicted in Fig. \ref{fig:01a} - note that the right panels depict the evolution of the Z-Score of the $S_{VN}$, as calculated against ensembles of equivalent random networks. 
It is clear that, even though the von Neumann entropy may be an alternative metric for comparing network structures and has a substantially lower computational cost, our proposed methodology is able to detect more complex changes, and is therefore more reliable in real-world situations.

\section{Application to real-world systems}
\label{sec:Applications}

\subsection{World-wide air transport network}
\label{sec:ATN}

As a first test case, we here consider the network created by flights between the top-50 and top-200 world airports, as extracted from the {\it Sabre Airport Data Intelligence} data set. As previously proposed \cite{sun2017worldwide, sun2017node}, nodes represent airports, pairwise connected when the total number of passengers per month who used a direct flight between both airports is larger than $1000$, {\it i.e.} at least $\approx 33$ passengers $/$ day. $72$ snapshots are available, representing the monthly evolution of the system between January 2010 and December 2015. 

\begin{figure*}[!tb]
\begin{center}
\includegraphics[width=0.8\textwidth]{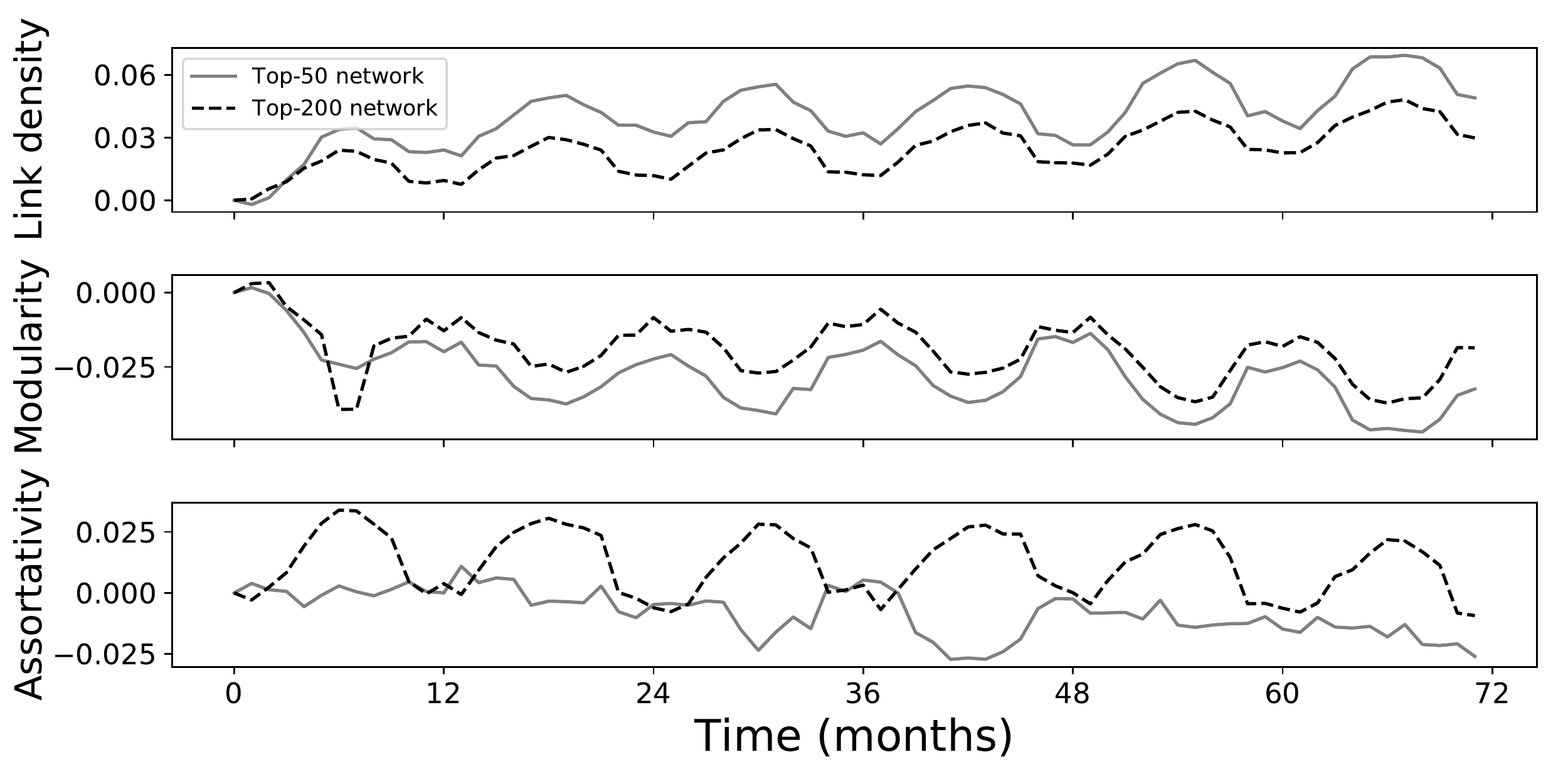}
\caption{Evolution of the link density, modularity (as calculated through the Louvain's algorithm \cite{blondel2008fast}) and assortativity of the world air transport network through time. Solid and dashed lines respectively represent the networks for the top-50 and top-200 world airports. All series are normalised by subtracting the value at $t=0$, in order to make them start from $0.0$. }\label{fig:02}
\end{center}
\end{figure*}

The air transport network is known to present a strong seasonality, both on the short ({\it i.e.} daily) and long scales (monthly and yearly) \cite{zanin2013modelling}. This magnifies the importance of using a correct temporal representation, as projecting the system into a single atemporal network may result in severe topological distortions \cite{rocha2017dynamics}. This fact is here confirmed by Fig. \ref{fig:02}, which represents the evolution of three topological metrics (link density, modularity and assortativity) through time - note the annual sinusoidal behaviour of all curves. For a detailed discussion of the effect of including different sets of airports on the observed topological metrics, the reader can refer to \cite{belkoura2016multi}.

\begin{figure*}[!tb]
\begin{center}
\includegraphics[width=0.9\textwidth]{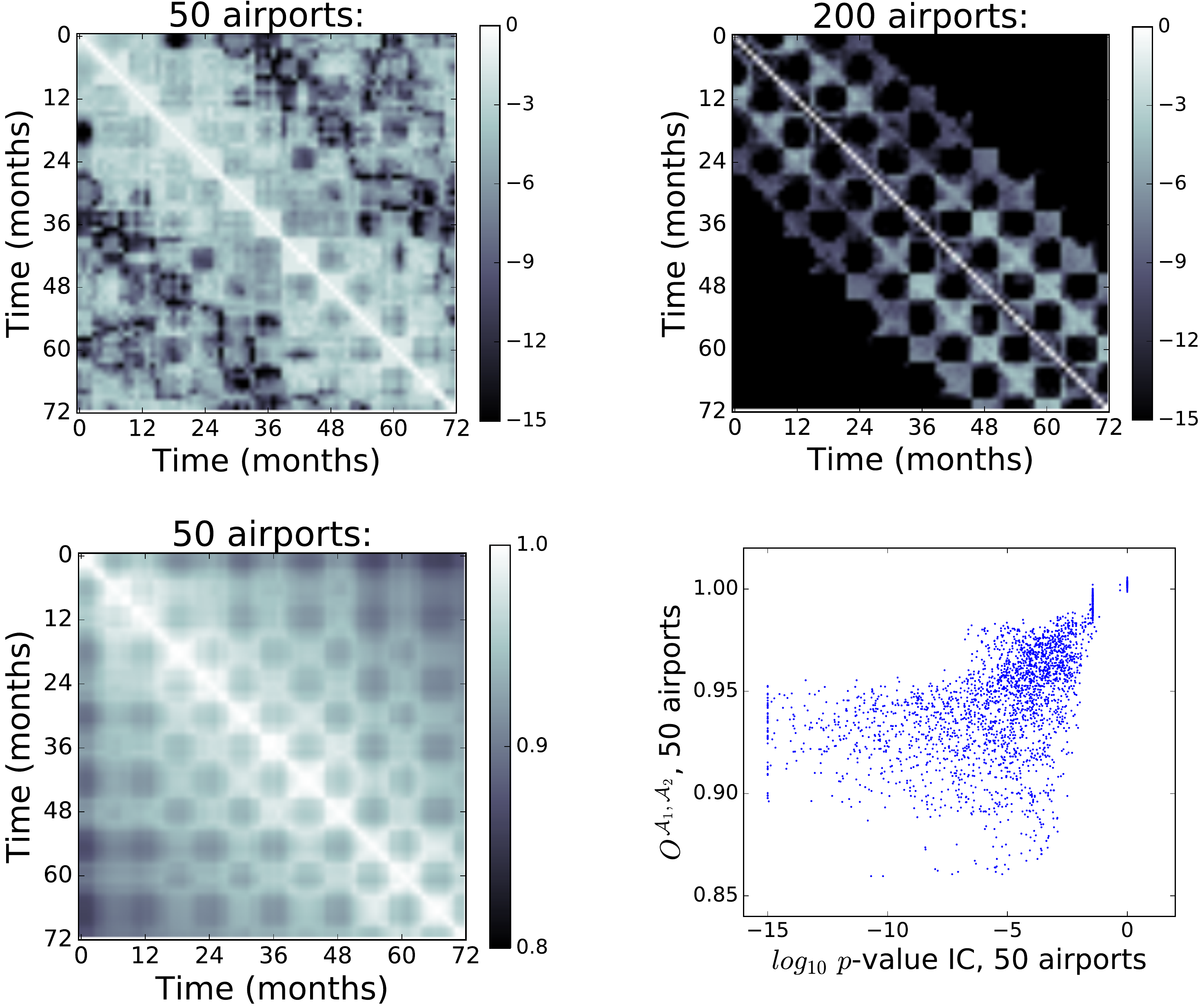}
\caption{Analysis of the world-wide air transport network. (Top) Evolution of the structure of changes, for the top-$50$ (left panel) and top-$200$ (right panel) networks. The colour of each point represents the $\log_{10}$ of the $p$-value of the $IC$ test, with light (dark) shades indicating random (structured) changes. (Bottom Left) Evolution of the structure, for the top-$50$ networks, as yielded by the normalised global overlap $O$. (Bottom Right) Scatter plot comparing the results yielded by the $IC$ and $O$, for the top-$50$ networks.}\label{fig:03}
\end{center}
\end{figure*}

The evolution of the $\log_{10}$ of the $p$-value of the $IC^\dagger$ test, for all possible pairs of months, is depicted in the top panels of Fig. \ref{fig:03}. Light colours represent changes with a random structure; dark colours the presence of a meso-scale regularity. In the case of $50$ airports, it is interesting to see bright squares on the main diagonal, of size $6 \times 6$, corresponding to the summer and winter seasons - this is to be expected, as flights seldom change within the same season, and differences are thus the consequence of small and random adjustments in the schedules. The yearly seasonality of the air transport is also evident in the case of $200$ airports, with bright colours concentrating around the $\pm 12$ and $\pm 24$ diagonals.
When the time distance between two snapshots is greater than two years, and when consecutive summer/winter pairs are compared, the $IC^\dagger$ test suggests that changes are not random: they thus correspond to systematic reconfigurations of the air transport market, driven by business considerations, and which cannot just be explained by a random rewiring.

As a comparison, Fig. \ref{fig:03} bottom left panel depicts the evolution of the normalised global overlap $O$ for the case of $50$ airports. While {\it prima facie} the colour map is similar to the one presented in Fig. \ref{fig:03} top left, several differences can be observed, especially far away from the main diagonal - {\it i.e.} for distances greater than 2 years. In order to clarify these differences, the bottom right panel reports a scatter plot comparing the values yielded by $IC$ and $O$. While there is a general positive correlation, it is possible to find completely different $IC$ values for a same overlap. For instance, for $O \approx 0.95$, one can find instances with $-15 < \log_{10} p$-value $ < -2$. This suggests that small changes, {\it i.e.} high overlaps, can be due to both (almost) random and strongly structured evolutions. The $IC$ thus yields a more complete view of the evolution of the network, providing information (specifically, the nature of the changes) that is disregard by other metrics.

\subsection{Hospital contact network}
\label{sec:Hosp}

As a second example, we here consider the temporal network of contacts in the geriatric unit of a Lyon university hospital, including patients and health care workers, as described in Refs. \cite{vanhems2013estimating,Lyon2017}. Nodes represent 46 health care workers and 29 patients, and links close-range interactions between them as detected by wearable sensors. The full data set spans from Monday, December 6, 2010 at 1:00 pm to Friday, December 10, 2010 at 2:00 pm, with a temporal resolution of 20 seconds. We extracted a set of 97 contact networks, by aggregating all contacts made within an one-hour interval, in order to avoid the sparsity characterising higher temporal resolutions. The median density of the network is $0.23$, the median and standard deviation of the average shortest path length is $2.07 \pm 0.38$, and the median clustering coefficient is $0.46$.

\begin{figure*}[!tb]
\begin{center}
\includegraphics[width=0.9\textwidth]{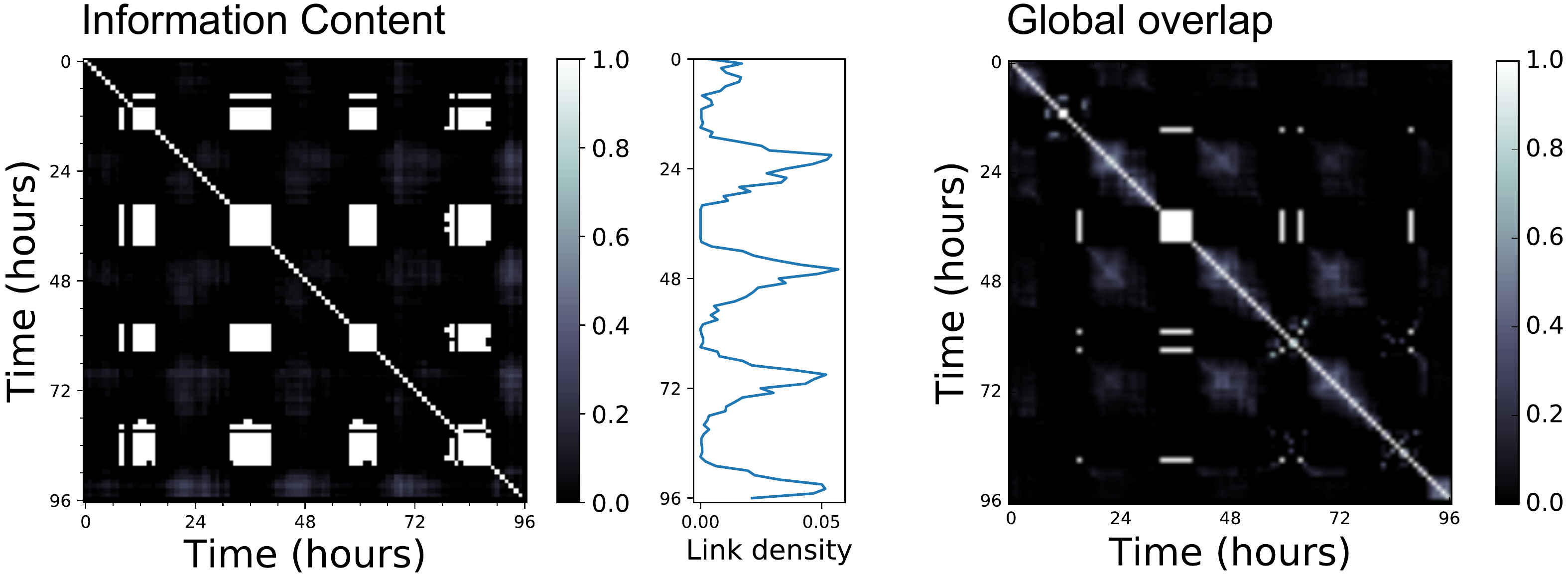}
\caption{Analysis of the hospital contact networks. (Left) Evolution of the structure of changes; note that the represented values correspond to $IC^*$. (Centre) Hourly evolution of the link density through the data set. (Right) Evolution of the structure as yielded by the normalised global overlap $O$.}\label{fig:04}
\end{center}
\end{figure*}

In a way similar to Fig. \ref{fig:03}, Fig. \ref{fig:04} (Left) represents the evolution of the structure of changes, for all pairs of available networks. Note that, in this case, $IC^*$ is used, such that values close to one (smaller than one) indicate random (respectively, structured) changes. A clear trend, with a $24$-hours period, can be identified - as confirmed by the central panel, depicting the evolution of the link density across several days.
A comparison between $IC$ and the global overlap can be made by considering the right panel of Fig. \ref{fig:04}, representing an equivalent analysis performed with the latter metric. Some interesting situations can be detected. For instance, one can observe that several time windows correspond to a very high $IC$ and, at the same time, to a very low global overlap - see, for instance, the bottom lower part of both colour maps. A high $IC$ can nevertheless also be found in the square in the main diagonal, at around $36$ hours, which corresponds to a high global overlap. The presence of a random change between two snapshots is thus not correlated with their overlap: it can appear when both a few or most of the links are rewired.

More in general, from the analysis of this system it can be concluded that it presents two different regimes. On one hand, most of the times the contact network evolves in a structured manner, reflecting the fact that health care workers perform regular tasks. On the other hand, nights are characterised by fewer contacts, which develop in a random fashion - possibly the result of emergencies and other random situations.

\subsection{Brain functional networks}
\label{sec:brain}

As a third case study, we present an analysis of the brain activity of multiple healthy subjects during a resting state, as made available by the Human Connectome Project (HCP) \cite{larson2013adding}.
Magnetoencephalographic (MEG) recordings \cite{hamalainen1993magnetoencephalography} were performed on a group of $10$ individuals, obtaining for each of them $248$ time series (each representing one MEG sensor) with $149,646$ points. Note that only a subset of the original group of people has been considered here, in order to ensure homogeneity in the number of channels and time series length.
Functional networks were then reconstructed as described in \cite{buldu2017frequency}, by firstly extracting the time series corresponding to four standard bands (theta $[3-8]$ Hz, alpha $[8-12]$ Hz, beta $[12-30]$ Hz, and gamma $[30-100]$ Hz); secondly, by calculating the Mutual Information (MI) between each pair of channels; and finally, by binarising the resulting networks, through a threshold defined by surrogated time series obtained by a block-permutation procedure \cite{canolty2006high}. The final result is thus a set of four functional networks per subject, representing brain activity at rest in four frequency bands. For further details about the recording and data processing, the reader is referred to Refs. \cite{larson2013adding, buldu2017frequency}.

As a first objective, we here want to show that the proposed algorithm can be used to quantify and describe the nature of the differences between the networks representing different frequency bands.
The average and standard deviation of $IC^*$ when comparing each person's four networks is reported in Tab. \ref{tab:Meg}. It can be seen that results are consistently included within the range $(0.78, 0.90)$, thus indicating the presence of structural differences between the networks corresponding to different frequency bands. This is to be expected, as these bands are supposed to correspond to different functional tasks, contributing differently to the overall resting state activity, and therefore not to be equivalent \cite{brookes2011investigating, hillebrand2012frequency}.

\begin{table*}[!tb]
\centering
\begin{tabular}{|l|c|c|c|c|}
\hline
 & {\bf Theta} & {\bf Alpha} & {\bf Beta} & {\bf Gamma} \\ \hline

{\bf Theta} & --- & $0.8111 \pm 0.1215$ & $0.8103 \pm 0.1099$ & $0.8943 \pm 0.0446$ \\ \hline

{\bf Alpha} & $0.8111 \pm 0.1215$ & --- & $0.7864 \pm 0.1555$ & $0.8166 \pm 0.1041$ \\ \hline

{\bf Beta} & $0.8103 \pm 0.1099$ & $0.7864 \pm 0.1555$ & --- & $0.8069 \pm 0.1254$ \\ \hline

{\bf Gamma} & $0.8943 \pm 0.0446$ & $0.8166 \pm 0.1041$ & $0.8069 \pm 0.1254$ & --- \\ \hline

\end{tabular}
\caption{Average and standard deviation of the $IC^*$ for the ten considered people, when networks of different frequency bands are pair-wise compared.}
\label{tab:Meg}
\end{table*}

\begin{figure*}[!tb]
\begin{center}
\includegraphics[width=0.90\textwidth]{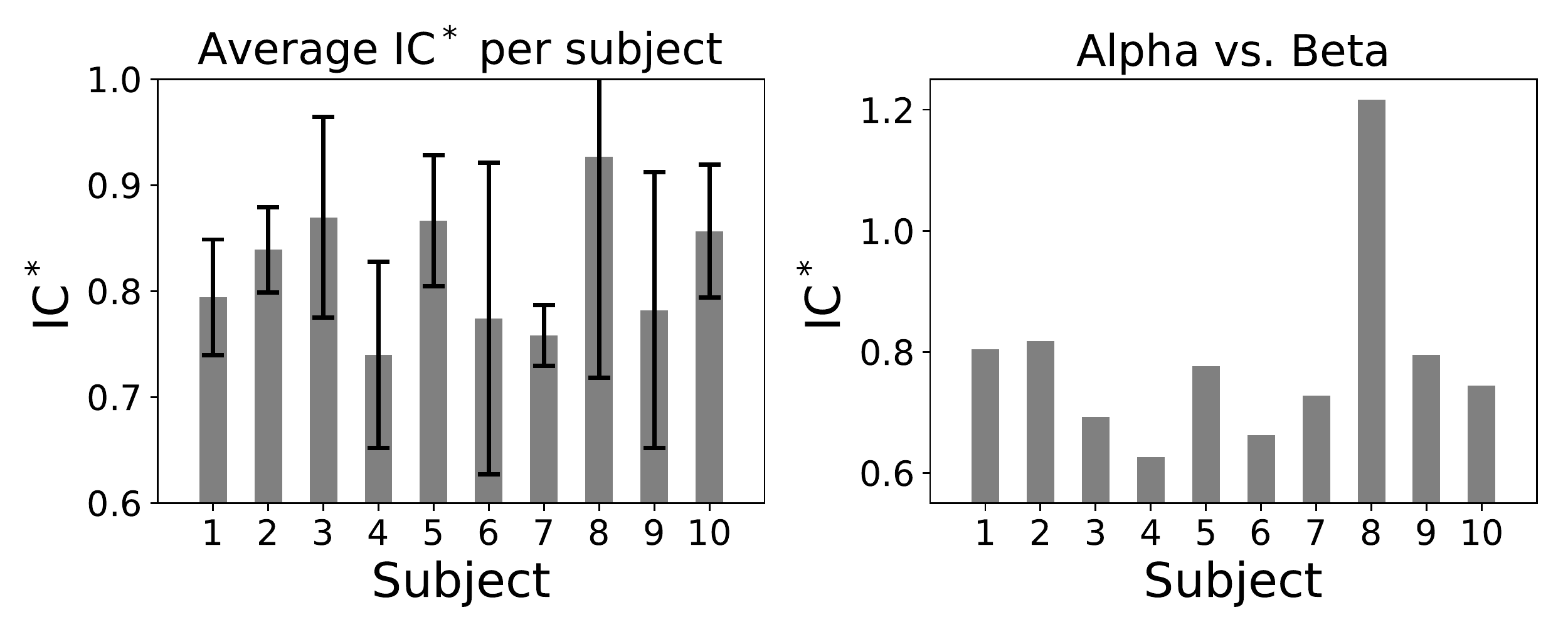}
\caption{(Left) Average and standard deviation of the $IC^*$ values for all pair-wise comparisons of the four (theta, alpha, beta and gamma) networks, as a function of the considered individual. (Right) $IC^*$ resulting of comparing the networks for the alpha and beta frequency bands, for all ten individuals.}\label{fig:05}
\end{center}
\end{figure*}

A quite different picture nevertheless arises when one shifts the focus to subjects. Fig. \ref{fig:05} (Left) depicts the average and standard deviation of the $IC^*$ values for each subject, {\it i.e.} corresponding to pair-wise comparing the four networks of each subject. A greater inter-subject variability emerges, with the average $IC^*$ varying between $0.740$ for subject $4$ and $0.927$ for subject $8$. An even stronger effect can be observed for the $IC^*$ between frequency bands alpha and beta, as depicted in Fig. \ref{fig:05} (Right): the same two subjects present values of respectively $0.627$ and $1.217$.

This last result highlights an important fact: alpha and beta bands can contribute to the global resting state activity in very different ways. In some of them, as in subject $4$, they have completely different topologies, while in others (as for subject $8$) their differences is only due to random fluctuations. More generally, different frequency bands interact between them in a way that is dependent on the subject, thus with high inter-subject variability. These results are aligned with previous findings, in which MEG studies report a low reproducibility for resting states in test-retest experiments \cite{deuker2009reproducibility, telesford2013exploration}.

\section{Finding a system's natural frequency: network self-correlations and correlograms}
\label{sec:Freq}

If a set of networks represents the evolution of the connectivity of a system through time, the parallelism with time series analysis can be pushed one step further by defining the equivalent of a network auto-correlation function. This requires calculating the similarity of the sequence of networks with itself, when one of the two instances is time-displaced with respect to the other.

Let us denote by $S$ the matrix of similarity, whose element $s_{i, j}$ encodes the similarity of the two networks respectively representing the system at times $i$ and $j$ - note that such matrix is completely equivalent to the results presented in Figs. \ref{fig:03} and \ref{fig:04}. The auto-correlation of the sequence of $N$ networks, for a time displacement of $t > 0$, is given by:

\begin{equation}
	C(t) = \frac{1}{N - t} \sum _{i = 0} ^{N - t} s_{i + t, i} = \frac{1}{N - t} \sum _{i = 0} ^{N - t} IC( A_{i+t}, A_i ).
\label{eq:self-corr}
\end{equation}

In the r.h.s. of Eq. \ref{eq:self-corr}, the $IC$ measure is used as a proxy of the similarity between two networks; to be more precise, this self-correlation thus assesses how a sequence of networks is {\it intentionally} equivalent to itself, excluding the presence of uncorrelated noise (unintentional changes) in the links. $C(t)$ is, by construction, equivalent to the average of the $t$-diagonal of $S$, or of the matrices depicted in Figs. \ref{fig:03} and \ref{fig:04}.

\begin{figure*}[!tb]
\begin{center}
\includegraphics[width=0.90\textwidth]{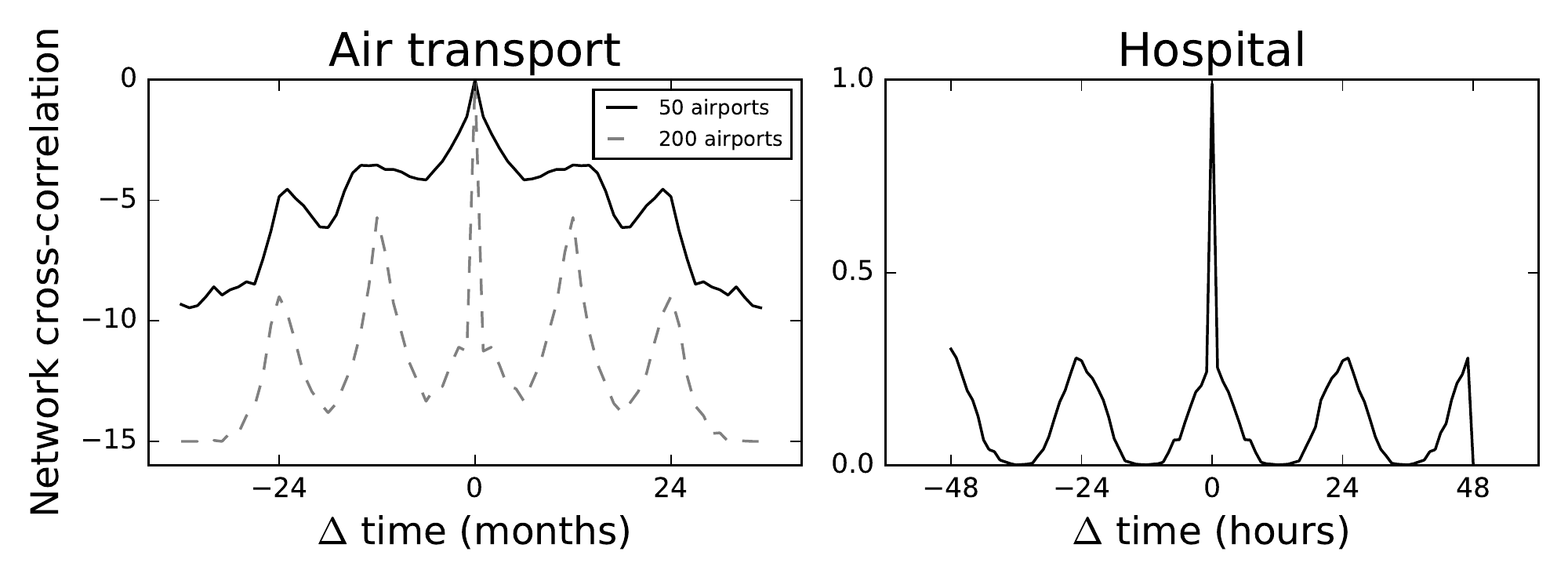}
\caption{Correlograms for the air transport (left panel) and hospital networks (right panel). While the Y axis of both panels represent the same concept, the specific measure changes: the $log_{10}$ of the $p$-value of $IC^\dagger$ in the former case, and $IC^*$ in the latter. See main text for definitions.}\label{fig:06}
\end{center}
\end{figure*}

By calculating $C(t)$ over all $t$s, it is possible to construct a full correlogram of the evolution of the studied system, with the maxima representing its natural frequencies.
In order to illustrate this idea, Fig. \ref{fig:06} depicts the correlograms for the air transport networks (left panel) and the hospital networks (right panel) - the brain functional networks have not here been considered, as they do not represent a temporal evolution. The respective matrices $S$ encode different variants of the $IC$ metric: the $log_{10}$ of the $p$-value of $IC^\dagger$ for the former (Fig. \ref{fig:04}), and $IC^*$ for the latter (Fig. \ref{fig:03}); as a consequence, the $y$ axis of the two panels have different scales. This is not a problem as long as the meaning is similar; in this case, both $IC^* \rightarrow 1$ and $IC^\dagger \rightarrow 0$ indicate highly similar networks, and both lie in the top part of the graph.
As should be expected, the maximum in both correlograms is located at $t = 0$. Local minima can additionally be found at $24k$ ($k \in \mathcal{Z}$) for the hospital data set, corresponding to a daily activity cycle; and at $12k$ ($k \in \mathcal{Z}$) in the case of the air transport, indicating a yearly seasonality.

As a final issue, it has to be highlighted that any other metric can be used within Eq. \ref{eq:self-corr} to calculate this correlogram, as for instance the global overlap $O$. Results would nevertheless have a different meaning, as the $IC$ allows not just to quantify the time required for returning to a given configuration, but also to ensure that differences are not due to structural changes.


\section{Discussion and conclusions}
\label{sec:Concl}

Beyond the assessment of the raw differences between two networks, a more complex and challenging problem is to detect if these differences are due to random modifications or to organised forces. The two problems are complementary and not necessarily correlated. The network structure of a system may substantially change between two measurements, but still be the same topology deformed by strong observational noise. On the other hand, small changes may be due to the targeted (intentional) attempt of, {\it e.g.}, promoting a node.

In this contribution we presented the use of Information Content \cite{zanin2014information} as a way of assessing the presence of meso-scale structures in the difference between two networks. The effectiveness of the metric has been demonstrated in several synthetic network evolutions, and tested with three real data sets respectively representing social, technological and biological systems. We additionally discussed the differences between the proposed approach and two {\it a priori} similar metrics, {\it i.e.} the network correlation \cite{bianconi2013statistical} and the von Neumann entropy \cite{braunstein2006laplacian, braunstein2006some}.

The availability of a similarity metric further allows to adapt some standard techniques in time series analysis to the study of the evolution of networked systems. We here considered the case of self-correlations and correlograms, and showed that the natural frequency of the system, in terms of recurrence of intentional network changes, can be estimated by the maxima in the network self-correlation. While not explicitly discussed here, the proposed analysis can be extended to the more general case of the cross-correlation, in which multiple sequences of networks, for instance representing two or more systems, can be pair-wise analysed. Correlograms could also be used to select the best time resolution for sampling temporal networks, a topic still to be explored \cite{rocha2017sampling}.

As a final thought, an hidden assumption of this work is that the networks to be compared are expected to be topologically compatible, {\it i.e.} to have the same number of nodes. While this holds for multiplex networks, general multi-layer and temporal graphs can have a variable size. The proposed methodology can still be used, provided an initial pre-processing is performed: for instance, the cores composed of nodes common to both networks could be isolated; while some information would be lost, the main evolutive trends could still be characterised.
Furthermore, networks coming from different systems, {\it e.g.} respectively representing brain activity and air transport, could in principle be compared. Nevertheless, it would firstly be necessary to {\it match} nodes between both networks, that is, to create a map relating each node of the first network with the topological equivalent one of the second, by means of {\it e.g.} the SimRank \cite{jeh2002simrank} or similar algorithms.


\bibliography{ComparingNets}

\providecommand{\noopsort}[1]{}\providecommand{\singleletter}[1]{#1}%
\begin{thebibliography}{40}%
\makeatletter
\providecommand \@ifxundefined [1]{%
 \@ifx{#1\undefined}
}%
\providecommand \@ifnum [1]{%
 \ifnum #1\expandafter \@firstoftwo
 \else \expandafter \@secondoftwo
 \fi
}%
\providecommand \@ifx [1]{%
 \ifx #1\expandafter \@firstoftwo
 \else \expandafter \@secondoftwo
 \fi
}%
\providecommand \natexlab [1]{#1}%
\providecommand \enquote  [1]{``#1''}%
\providecommand \bibnamefont  [1]{#1}%
\providecommand \bibfnamefont [1]{#1}%
\providecommand \citenamefont [1]{#1}%
\providecommand \href@noop [0]{\@secondoftwo}%
\providecommand \href [0]{\begingroup \@sanitize@url \@href}%
\providecommand \@href[1]{\@@startlink{#1}\@@href}%
\providecommand \@@href[1]{\endgroup#1\@@endlink}%
\providecommand \@sanitize@url [0]{\catcode `\\12\catcode `\$12\catcode
  `\&12\catcode `\#12\catcode `\^12\catcode `\_12\catcode `\%12\relax}%
\providecommand \@@startlink[1]{}%
\providecommand \@@endlink[0]{}%
\providecommand \url  [0]{\begingroup\@sanitize@url \@url }%
\providecommand \@url [1]{\endgroup\@href {#1}{\urlprefix }}%
\providecommand \urlprefix  [0]{URL }%
\providecommand \Eprint [0]{\href }%
\providecommand \doibase [0]{http://dx.doi.org/}%
\providecommand \selectlanguage [0]{\@gobble}%
\providecommand \bibinfo  [0]{\@secondoftwo}%
\providecommand \bibfield  [0]{\@secondoftwo}%
\providecommand \translation [1]{[#1]}%
\providecommand \BibitemOpen [0]{}%
\providecommand \bibitemStop [0]{}%
\providecommand \bibitemNoStop [0]{.\EOS\space}%
\providecommand \EOS [0]{\spacefactor3000\relax}%
\providecommand \BibitemShut  [1]{\csname bibitem#1\endcsname}%
\let\auto@bib@innerbib\@empty
\bibitem [{\citenamefont {Strogatz}(2001)}]{strogatz2001exploring}%
  \BibitemOpen
  \bibfield  {author} {\bibinfo {author} {\bibfnamefont {S.~H.}\ \bibnamefont
  {Strogatz}},\ }\href@noop {} {\bibfield  {journal} {\bibinfo  {journal}
  {Nature}\ }\textbf {\bibinfo {volume} {410}},\ \bibinfo {pages} {268}
  (\bibinfo {year} {2001})}\BibitemShut {NoStop}%
\bibitem [{\citenamefont {Newman}(2003)}]{newman2003structure}%
  \BibitemOpen
  \bibfield  {author} {\bibinfo {author} {\bibfnamefont {M.~E.}\ \bibnamefont
  {Newman}},\ }\href@noop {} {\bibfield  {journal} {\bibinfo  {journal} {SIAM
  review}\ }\textbf {\bibinfo {volume} {45}},\ \bibinfo {pages} {167} (\bibinfo
  {year} {2003})}\BibitemShut {NoStop}%
\bibitem [{\citenamefont {Greicius}\ \emph {et~al.}(2003)\citenamefont
  {Greicius}, \citenamefont {Krasnow}, \citenamefont {Reiss},\ and\
  \citenamefont {Menon}}]{greicius2003functional}%
  \BibitemOpen
  \bibfield  {author} {\bibinfo {author} {\bibfnamefont {M.~D.}\ \bibnamefont
  {Greicius}}, \bibinfo {author} {\bibfnamefont {B.}~\bibnamefont {Krasnow}},
  \bibinfo {author} {\bibfnamefont {A.~L.}\ \bibnamefont {Reiss}}, \ and\
  \bibinfo {author} {\bibfnamefont {V.}~\bibnamefont {Menon}},\ }\href@noop {}
  {\bibfield  {journal} {\bibinfo  {journal} {Proceedings of the National
  Academy of Sciences}\ }\textbf {\bibinfo {volume} {100}},\ \bibinfo {pages}
  {253} (\bibinfo {year} {2003})}\BibitemShut {NoStop}%
\bibitem [{\citenamefont {Seeley}\ \emph {et~al.}(2009)\citenamefont {Seeley},
  \citenamefont {Crawford}, \citenamefont {Zhou}, \citenamefont {Miller},\ and\
  \citenamefont {Greicius}}]{seeley2009neurodegenerative}%
  \BibitemOpen
  \bibfield  {author} {\bibinfo {author} {\bibfnamefont {W.~W.}\ \bibnamefont
  {Seeley}}, \bibinfo {author} {\bibfnamefont {R.~K.}\ \bibnamefont
  {Crawford}}, \bibinfo {author} {\bibfnamefont {J.}~\bibnamefont {Zhou}},
  \bibinfo {author} {\bibfnamefont {B.~L.}\ \bibnamefont {Miller}}, \ and\
  \bibinfo {author} {\bibfnamefont {M.~D.}\ \bibnamefont {Greicius}},\
  }\href@noop {} {\bibfield  {journal} {\bibinfo  {journal} {Neuron}\ }\textbf
  {\bibinfo {volume} {62}},\ \bibinfo {pages} {42} (\bibinfo {year}
  {2009})}\BibitemShut {NoStop}%
\bibitem [{\citenamefont {Bassett}\ \emph {et~al.}(2011)\citenamefont
  {Bassett}, \citenamefont {Wymbs}, \citenamefont {Porter}, \citenamefont
  {Mucha}, \citenamefont {Carlson},\ and\ \citenamefont
  {Grafton}}]{bassett2011dynamic}%
  \BibitemOpen
  \bibfield  {author} {\bibinfo {author} {\bibfnamefont {D.~S.}\ \bibnamefont
  {Bassett}}, \bibinfo {author} {\bibfnamefont {N.~F.}\ \bibnamefont {Wymbs}},
  \bibinfo {author} {\bibfnamefont {M.~A.}\ \bibnamefont {Porter}}, \bibinfo
  {author} {\bibfnamefont {P.~J.}\ \bibnamefont {Mucha}}, \bibinfo {author}
  {\bibfnamefont {J.~M.}\ \bibnamefont {Carlson}}, \ and\ \bibinfo {author}
  {\bibfnamefont {S.~T.}\ \bibnamefont {Grafton}},\ }\href@noop {} {\bibfield
  {journal} {\bibinfo  {journal} {Proceedings of the National Academy of
  Sciences}\ }\textbf {\bibinfo {volume} {108}},\ \bibinfo {pages} {7641}
  (\bibinfo {year} {2011})}\BibitemShut {NoStop}%
\bibitem [{\citenamefont {Zalesky}\ \emph {et~al.}(2014)\citenamefont
  {Zalesky}, \citenamefont {Fornito}, \citenamefont {Cocchi}, \citenamefont
  {Gollo},\ and\ \citenamefont {Breakspear}}]{zalesky2014time}%
  \BibitemOpen
  \bibfield  {author} {\bibinfo {author} {\bibfnamefont {A.}~\bibnamefont
  {Zalesky}}, \bibinfo {author} {\bibfnamefont {A.}~\bibnamefont {Fornito}},
  \bibinfo {author} {\bibfnamefont {L.}~\bibnamefont {Cocchi}}, \bibinfo
  {author} {\bibfnamefont {L.~L.}\ \bibnamefont {Gollo}}, \ and\ \bibinfo
  {author} {\bibfnamefont {M.}~\bibnamefont {Breakspear}},\ }\href@noop {}
  {\bibfield  {journal} {\bibinfo  {journal} {Proceedings of the National
  Academy of Sciences}\ }\textbf {\bibinfo {volume} {111}},\ \bibinfo {pages}
  {10341} (\bibinfo {year} {2014})}\BibitemShut {NoStop}%
\bibitem [{\citenamefont {Buld{\'u}}\ and\ \citenamefont
  {Porter}(2017)}]{buldu2017frequency}%
  \BibitemOpen
  \bibfield  {author} {\bibinfo {author} {\bibfnamefont {J.~M.}\ \bibnamefont
  {Buld{\'u}}}\ and\ \bibinfo {author} {\bibfnamefont {M.~A.}\ \bibnamefont
  {Porter}},\ }\href@noop {} {\bibfield  {journal} {\bibinfo  {journal} {arXiv
  preprint arXiv:1703.06091}\ } (\bibinfo {year} {2017})}\BibitemShut {NoStop}%
\bibitem [{\citenamefont {Costa}\ \emph {et~al.}(2011)\citenamefont {Costa},
  \citenamefont {Oliveira~Jr}, \citenamefont {Travieso}, \citenamefont
  {Rodrigues}, \citenamefont {Villas~Boas}, \citenamefont {Antiqueira},
  \citenamefont {Viana},\ and\ \citenamefont
  {Correa~Rocha}}]{costa2011analyzing}%
  \BibitemOpen
  \bibfield  {author} {\bibinfo {author} {\bibfnamefont {L.~d.~F.}\
  \bibnamefont {Costa}}, \bibinfo {author} {\bibfnamefont {O.~N.}\ \bibnamefont
  {Oliveira~Jr}}, \bibinfo {author} {\bibfnamefont {G.}~\bibnamefont
  {Travieso}}, \bibinfo {author} {\bibfnamefont {F.~A.}\ \bibnamefont
  {Rodrigues}}, \bibinfo {author} {\bibfnamefont {P.~R.}\ \bibnamefont
  {Villas~Boas}}, \bibinfo {author} {\bibfnamefont {L.}~\bibnamefont
  {Antiqueira}}, \bibinfo {author} {\bibfnamefont {M.~P.}\ \bibnamefont
  {Viana}}, \ and\ \bibinfo {author} {\bibfnamefont {L.~E.}\ \bibnamefont
  {Correa~Rocha}},\ }\href@noop {} {\bibfield  {journal} {\bibinfo  {journal}
  {Advances in Physics}\ }\textbf {\bibinfo {volume} {60}},\ \bibinfo {pages}
  {329} (\bibinfo {year} {2011})}\BibitemShut {NoStop}%
\bibitem [{\citenamefont {Neal}(2014)}]{neal2014devil}%
  \BibitemOpen
  \bibfield  {author} {\bibinfo {author} {\bibfnamefont {Z.}~\bibnamefont
  {Neal}},\ }\href@noop {} {\bibfield  {journal} {\bibinfo  {journal} {Social
  networks}\ }\textbf {\bibinfo {volume} {38}},\ \bibinfo {pages} {63}
  (\bibinfo {year} {2014})}\BibitemShut {NoStop}%
\bibitem [{\citenamefont {Belkoura}\ \emph {et~al.}(2016)\citenamefont
  {Belkoura}, \citenamefont {Cook}, \citenamefont {Pe{\~n}a},\ and\
  \citenamefont {Zanin}}]{belkoura2016multi}%
  \BibitemOpen
  \bibfield  {author} {\bibinfo {author} {\bibfnamefont {S.}~\bibnamefont
  {Belkoura}}, \bibinfo {author} {\bibfnamefont {A.}~\bibnamefont {Cook}},
  \bibinfo {author} {\bibfnamefont {J.~M.}\ \bibnamefont {Pe{\~n}a}}, \ and\
  \bibinfo {author} {\bibfnamefont {M.}~\bibnamefont {Zanin}},\ }\href@noop {}
  {\bibfield  {journal} {\bibinfo  {journal} {Transportation Research Part E:
  Logistics and Transportation Review}\ }\textbf {\bibinfo {volume} {94}},\
  \bibinfo {pages} {95} (\bibinfo {year} {2016})}\BibitemShut {NoStop}%
\bibitem [{\citenamefont {Br{\'o}dka}\ \emph {et~al.}(2017)\citenamefont
  {Br{\'o}dka}, \citenamefont {Chmiel}, \citenamefont {Magnani},\ and\
  \citenamefont {Ragozini}}]{brodka2017quantifying}%
  \BibitemOpen
  \bibfield  {author} {\bibinfo {author} {\bibfnamefont {P.}~\bibnamefont
  {Br{\'o}dka}}, \bibinfo {author} {\bibfnamefont {A.}~\bibnamefont {Chmiel}},
  \bibinfo {author} {\bibfnamefont {M.}~\bibnamefont {Magnani}}, \ and\
  \bibinfo {author} {\bibfnamefont {G.}~\bibnamefont {Ragozini}},\ }\href@noop
  {} {\bibfield  {journal} {\bibinfo  {journal} {arXiv preprint
  arXiv:1711.11335}\ } (\bibinfo {year} {2017})}\BibitemShut {NoStop}%
\bibitem [{\citenamefont {Costa}\ \emph {et~al.}(2007)\citenamefont {Costa},
  \citenamefont {Rodrigues}, \citenamefont {Travieso},\ and\ \citenamefont
  {Villas~Boas}}]{costa2007characterization}%
  \BibitemOpen
  \bibfield  {author} {\bibinfo {author} {\bibfnamefont {L.~d.~F.}\
  \bibnamefont {Costa}}, \bibinfo {author} {\bibfnamefont {F.~A.}\ \bibnamefont
  {Rodrigues}}, \bibinfo {author} {\bibfnamefont {G.}~\bibnamefont {Travieso}},
  \ and\ \bibinfo {author} {\bibfnamefont {P.~R.}\ \bibnamefont
  {Villas~Boas}},\ }\href@noop {} {\bibfield  {journal} {\bibinfo  {journal}
  {Advances in physics}\ }\textbf {\bibinfo {volume} {56}},\ \bibinfo {pages}
  {167} (\bibinfo {year} {2007})}\BibitemShut {NoStop}%
\bibitem [{\citenamefont {Strohman}(2002)}]{strohman2002maneuvering}%
  \BibitemOpen
  \bibfield  {author} {\bibinfo {author} {\bibfnamefont {R.}~\bibnamefont
  {Strohman}},\ }\href@noop {} {\bibfield  {journal} {\bibinfo  {journal}
  {Science}\ }\textbf {\bibinfo {volume} {296}},\ \bibinfo {pages} {701}
  (\bibinfo {year} {2002})}\BibitemShut {NoStop}%
\bibitem [{\citenamefont {Zanin}\ \emph {et~al.}(2016)\citenamefont {Zanin},
  \citenamefont {Correia}, \citenamefont {Sousa},\ and\ \citenamefont
  {Cruz}}]{zanin2016phenotype}%
  \BibitemOpen
  \bibfield  {author} {\bibinfo {author} {\bibfnamefont {M.}~\bibnamefont
  {Zanin}}, \bibinfo {author} {\bibfnamefont {M.}~\bibnamefont {Correia}},
  \bibinfo {author} {\bibfnamefont {P.~A.}\ \bibnamefont {Sousa}}, \ and\
  \bibinfo {author} {\bibfnamefont {J.}~\bibnamefont {Cruz}},\ }\href@noop {}
  {\bibfield  {journal} {\bibinfo  {journal} {Scientific reports}\ }\textbf
  {\bibinfo {volume} {6}},\ \bibinfo {pages} {19790} (\bibinfo {year}
  {2016})}\BibitemShut {NoStop}%
\bibitem [{\citenamefont {Zanin}, \citenamefont {Sousa},\ and\ \citenamefont
  {Menasalvas}(2014)}]{zanin2014information}%
  \BibitemOpen
  \bibfield  {author} {\bibinfo {author} {\bibfnamefont {M.}~\bibnamefont
  {Zanin}}, \bibinfo {author} {\bibfnamefont {P.~A.}\ \bibnamefont {Sousa}}, \
  and\ \bibinfo {author} {\bibfnamefont {E.}~\bibnamefont {Menasalvas}},\
  }\href@noop {} {\bibfield  {journal} {\bibinfo  {journal} {EPL (Europhysics
  letters)}\ }\textbf {\bibinfo {volume} {106}},\ \bibinfo {pages} {30001}
  (\bibinfo {year} {2014})}\BibitemShut {NoStop}%
\bibitem [{\citenamefont {Bianconi}(2013)}]{bianconi2013statistical}%
  \BibitemOpen
  \bibfield  {author} {\bibinfo {author} {\bibfnamefont {G.}~\bibnamefont
  {Bianconi}},\ }\href@noop {} {\bibfield  {journal} {\bibinfo  {journal}
  {Physical Review E}\ }\textbf {\bibinfo {volume} {87}},\ \bibinfo {pages}
  {062806} (\bibinfo {year} {2013})}\BibitemShut {NoStop}%
\bibitem [{\citenamefont {Braunstein}, \citenamefont {Ghosh},\ and\
  \citenamefont {Severini}(2006)}]{braunstein2006laplacian}%
  \BibitemOpen
  \bibfield  {author} {\bibinfo {author} {\bibfnamefont {S.~L.}\ \bibnamefont
  {Braunstein}}, \bibinfo {author} {\bibfnamefont {S.}~\bibnamefont {Ghosh}}, \
  and\ \bibinfo {author} {\bibfnamefont {S.}~\bibnamefont {Severini}},\
  }\href@noop {} {\bibfield  {journal} {\bibinfo  {journal} {Annals of
  Combinatorics}\ }\textbf {\bibinfo {volume} {10}},\ \bibinfo {pages} {291}
  (\bibinfo {year} {2006})}\BibitemShut {NoStop}%
\bibitem [{\citenamefont {Braunstein}\ \emph {et~al.}(2006)\citenamefont
  {Braunstein}, \citenamefont {Ghosh}, \citenamefont {Mansour}, \citenamefont
  {Severini},\ and\ \citenamefont {Wilson}}]{braunstein2006some}%
  \BibitemOpen
  \bibfield  {author} {\bibinfo {author} {\bibfnamefont {S.~L.}\ \bibnamefont
  {Braunstein}}, \bibinfo {author} {\bibfnamefont {S.}~\bibnamefont {Ghosh}},
  \bibinfo {author} {\bibfnamefont {T.}~\bibnamefont {Mansour}}, \bibinfo
  {author} {\bibfnamefont {S.}~\bibnamefont {Severini}}, \ and\ \bibinfo
  {author} {\bibfnamefont {R.~C.}\ \bibnamefont {Wilson}},\ }\href@noop {}
  {\bibfield  {journal} {\bibinfo  {journal} {Physical Review A}\ }\textbf
  {\bibinfo {volume} {73}},\ \bibinfo {pages} {012320} (\bibinfo {year}
  {2006})}\BibitemShut {NoStop}%
\bibitem [{\citenamefont {Boccaletti}\ \emph {et~al.}(2014)\citenamefont
  {Boccaletti}, \citenamefont {Bianconi}, \citenamefont {Criado}, \citenamefont
  {Del~Genio}, \citenamefont {G{\'o}mez-Gardenes}, \citenamefont {Romance},
  \citenamefont {Sendina-Nadal}, \citenamefont {Wang},\ and\ \citenamefont
  {Zanin}}]{boccaletti2014structure}%
  \BibitemOpen
  \bibfield  {author} {\bibinfo {author} {\bibfnamefont {S.}~\bibnamefont
  {Boccaletti}}, \bibinfo {author} {\bibfnamefont {G.}~\bibnamefont
  {Bianconi}}, \bibinfo {author} {\bibfnamefont {R.}~\bibnamefont {Criado}},
  \bibinfo {author} {\bibfnamefont {C.~I.}\ \bibnamefont {Del~Genio}}, \bibinfo
  {author} {\bibfnamefont {J.}~\bibnamefont {G{\'o}mez-Gardenes}}, \bibinfo
  {author} {\bibfnamefont {M.}~\bibnamefont {Romance}}, \bibinfo {author}
  {\bibfnamefont {I.}~\bibnamefont {Sendina-Nadal}}, \bibinfo {author}
  {\bibfnamefont {Z.}~\bibnamefont {Wang}}, \ and\ \bibinfo {author}
  {\bibfnamefont {M.}~\bibnamefont {Zanin}},\ }\href@noop {} {\bibfield
  {journal} {\bibinfo  {journal} {Physics Reports}\ }\textbf {\bibinfo {volume}
  {544}},\ \bibinfo {pages} {1} (\bibinfo {year} {2014})}\BibitemShut {NoStop}%
\bibitem [{\citenamefont {Holme}\ and\ \citenamefont
  {Saram{\"a}ki}(2012)}]{holme2012temporal}%
  \BibitemOpen
  \bibfield  {author} {\bibinfo {author} {\bibfnamefont {P.}~\bibnamefont
  {Holme}}\ and\ \bibinfo {author} {\bibfnamefont {J.}~\bibnamefont
  {Saram{\"a}ki}},\ }\href@noop {} {\bibfield  {journal} {\bibinfo  {journal}
  {Physics reports}\ }\textbf {\bibinfo {volume} {519}},\ \bibinfo {pages} {97}
  (\bibinfo {year} {2012})}\BibitemShut {NoStop}%
\bibitem [{\citenamefont {Cellai}\ \emph {et~al.}(2013)\citenamefont {Cellai},
  \citenamefont {L{\'o}pez}, \citenamefont {Zhou}, \citenamefont {Gleeson},\
  and\ \citenamefont {Bianconi}}]{cellai2013percolation}%
  \BibitemOpen
  \bibfield  {author} {\bibinfo {author} {\bibfnamefont {D.}~\bibnamefont
  {Cellai}}, \bibinfo {author} {\bibfnamefont {E.}~\bibnamefont {L{\'o}pez}},
  \bibinfo {author} {\bibfnamefont {J.}~\bibnamefont {Zhou}}, \bibinfo {author}
  {\bibfnamefont {J.~P.}\ \bibnamefont {Gleeson}}, \ and\ \bibinfo {author}
  {\bibfnamefont {G.}~\bibnamefont {Bianconi}},\ }\href@noop {} {\bibfield
  {journal} {\bibinfo  {journal} {Physical Review E}\ }\textbf {\bibinfo
  {volume} {88}},\ \bibinfo {pages} {052811} (\bibinfo {year}
  {2013})}\BibitemShut {NoStop}%
\bibitem [{\citenamefont {Baxter}\ \emph {et~al.}(2016)\citenamefont {Baxter},
  \citenamefont {Bianconi}, \citenamefont {da~Costa}, \citenamefont
  {Dorogovtsev},\ and\ \citenamefont {Mendes}}]{baxter2016correlated}%
  \BibitemOpen
  \bibfield  {author} {\bibinfo {author} {\bibfnamefont {G.~J.}\ \bibnamefont
  {Baxter}}, \bibinfo {author} {\bibfnamefont {G.}~\bibnamefont {Bianconi}},
  \bibinfo {author} {\bibfnamefont {R.~A.}\ \bibnamefont {da~Costa}}, \bibinfo
  {author} {\bibfnamefont {S.~N.}\ \bibnamefont {Dorogovtsev}}, \ and\ \bibinfo
  {author} {\bibfnamefont {J.~F.~F.}\ \bibnamefont {Mendes}},\ }\href@noop {}
  {\bibfield  {journal} {\bibinfo  {journal} {Physical Review E}\ }\textbf
  {\bibinfo {volume} {94}},\ \bibinfo {pages} {012303} (\bibinfo {year}
  {2016})}\BibitemShut {NoStop}%
\bibitem [{\citenamefont {Passerini}\ and\ \citenamefont
  {Severini}(2009)}]{passerini2009quantifying}%
  \BibitemOpen
  \bibfield  {author} {\bibinfo {author} {\bibfnamefont {F.}~\bibnamefont
  {Passerini}}\ and\ \bibinfo {author} {\bibfnamefont {S.}~\bibnamefont
  {Severini}},\ }\href@noop {} {\bibfield  {journal} {\bibinfo  {journal}
  {International Journal of Agent Technologies and Systems (IJATS)}\ }\textbf
  {\bibinfo {volume} {1}},\ \bibinfo {pages} {58} (\bibinfo {year}
  {2009})}\BibitemShut {NoStop}%
\bibitem [{\citenamefont {De~Domenico}\ and\ \citenamefont
  {Biamonte}(2016)}]{de2016spectral}%
  \BibitemOpen
  \bibfield  {author} {\bibinfo {author} {\bibfnamefont {M.}~\bibnamefont
  {De~Domenico}}\ and\ \bibinfo {author} {\bibfnamefont {J.}~\bibnamefont
  {Biamonte}},\ }\href@noop {} {\bibfield  {journal} {\bibinfo  {journal}
  {Physical Review X}\ }\textbf {\bibinfo {volume} {6}},\ \bibinfo {pages}
  {041062} (\bibinfo {year} {2016})}\BibitemShut {NoStop}%
\bibitem [{\citenamefont {Sun}, \citenamefont {Wandelt},\ and\ \citenamefont
  {Zanin}(2017)}]{sun2017worldwide}%
  \BibitemOpen
  \bibfield  {author} {\bibinfo {author} {\bibfnamefont {X.}~\bibnamefont
  {Sun}}, \bibinfo {author} {\bibfnamefont {S.}~\bibnamefont {Wandelt}}, \ and\
  \bibinfo {author} {\bibfnamefont {M.}~\bibnamefont {Zanin}},\ }\href@noop {}
  {\bibfield  {journal} {\bibinfo  {journal} {Transportmetrica A: Transport
  Science}\ ,\ \bibinfo {pages} {1}} (\bibinfo {year} {2017})}\BibitemShut
  {NoStop}%
\bibitem [{\citenamefont {Sun}, \citenamefont {Wandelt},\ and\ \citenamefont
  {Cao}(2017)}]{sun2017node}%
  \BibitemOpen
  \bibfield  {author} {\bibinfo {author} {\bibfnamefont {X.}~\bibnamefont
  {Sun}}, \bibinfo {author} {\bibfnamefont {S.}~\bibnamefont {Wandelt}}, \ and\
  \bibinfo {author} {\bibfnamefont {X.}~\bibnamefont {Cao}},\ }\href@noop {}
  {\bibfield  {journal} {\bibinfo  {journal} {Networks and Spatial Economics}\
  ,\ \bibinfo {pages} {1}} (\bibinfo {year} {2017})}\BibitemShut {NoStop}%
\bibitem [{\citenamefont {Zanin}\ and\ \citenamefont
  {Lillo}(2013)}]{zanin2013modelling}%
  \BibitemOpen
  \bibfield  {author} {\bibinfo {author} {\bibfnamefont {M.}~\bibnamefont
  {Zanin}}\ and\ \bibinfo {author} {\bibfnamefont {F.}~\bibnamefont {Lillo}},\
  }\href@noop {} {\bibfield  {journal} {\bibinfo  {journal} {The European
  Physical Journal Special Topics}\ }\textbf {\bibinfo {volume} {215}},\
  \bibinfo {pages} {5} (\bibinfo {year} {2013})}\BibitemShut {NoStop}%
\bibitem [{\citenamefont {Rocha}(2017)}]{rocha2017dynamics}%
  \BibitemOpen
  \bibfield  {author} {\bibinfo {author} {\bibfnamefont {L.~E.~C.}\
  \bibnamefont {Rocha}},\ }\href@noop {} {\bibfield  {journal} {\bibinfo
  {journal} {Chinese Journal of Aeronautics}\ }\textbf {\bibinfo {volume}
  {30}},\ \bibinfo {pages} {469} (\bibinfo {year} {2017})}\BibitemShut
  {NoStop}%
\bibitem [{\citenamefont {Blondel}\ \emph {et~al.}(2008)\citenamefont
  {Blondel}, \citenamefont {Guillaume}, \citenamefont {Lambiotte},\ and\
  \citenamefont {Lefebvre}}]{blondel2008fast}%
  \BibitemOpen
  \bibfield  {author} {\bibinfo {author} {\bibfnamefont {V.~D.}\ \bibnamefont
  {Blondel}}, \bibinfo {author} {\bibfnamefont {J.-L.}\ \bibnamefont
  {Guillaume}}, \bibinfo {author} {\bibfnamefont {R.}~\bibnamefont
  {Lambiotte}}, \ and\ \bibinfo {author} {\bibfnamefont {E.}~\bibnamefont
  {Lefebvre}},\ }\href@noop {} {\bibfield  {journal} {\bibinfo  {journal}
  {Journal of statistical mechanics: theory and experiment}\ }\textbf {\bibinfo
  {volume} {2008}},\ \bibinfo {pages} {P10008} (\bibinfo {year}
  {2008})}\BibitemShut {NoStop}%
\bibitem [{\citenamefont {Vanhems}\ \emph {et~al.}(2013)\citenamefont
  {Vanhems}, \citenamefont {Barrat}, \citenamefont {Cattuto}, \citenamefont
  {Pinton}, \citenamefont {Khanafer}, \citenamefont {R{\'e}gis}, \citenamefont
  {Kim}, \citenamefont {Comte},\ and\ \citenamefont
  {Voirin}}]{vanhems2013estimating}%
  \BibitemOpen
  \bibfield  {author} {\bibinfo {author} {\bibfnamefont {P.}~\bibnamefont
  {Vanhems}}, \bibinfo {author} {\bibfnamefont {A.}~\bibnamefont {Barrat}},
  \bibinfo {author} {\bibfnamefont {C.}~\bibnamefont {Cattuto}}, \bibinfo
  {author} {\bibfnamefont {J.-F.}\ \bibnamefont {Pinton}}, \bibinfo {author}
  {\bibfnamefont {N.}~\bibnamefont {Khanafer}}, \bibinfo {author}
  {\bibfnamefont {C.}~\bibnamefont {R{\'e}gis}}, \bibinfo {author}
  {\bibfnamefont {B.-a.}\ \bibnamefont {Kim}}, \bibinfo {author} {\bibfnamefont
  {B.}~\bibnamefont {Comte}}, \ and\ \bibinfo {author} {\bibfnamefont
  {N.}~\bibnamefont {Voirin}},\ }\href@noop {} {\bibfield  {journal} {\bibinfo
  {journal} {PloS one}\ }\textbf {\bibinfo {volume} {8}},\ \bibinfo {pages}
  {e73970} (\bibinfo {year} {2013})}\BibitemShut {NoStop}%
\bibitem [{Lyo()}]{Lyon2017}%
  \BibitemOpen
  \href@noop {} {\enquote {\bibinfo {title} {Sociopatterns hospital network},}\
  }\bibinfo {howpublished} {\url{http://www.sociopatterns.org}},\ \bibinfo
  {note} {accessed: 2017-09-01}\BibitemShut {NoStop}%
\bibitem [{\citenamefont {Larson-Prior}\ \emph {et~al.}(2013)\citenamefont
  {Larson-Prior}, \citenamefont {Oostenveld}, \citenamefont {Della~Penna},
  \citenamefont {Michalareas}, \citenamefont {Prior}, \citenamefont
  {Babajani-Feremi}, \citenamefont {Schoffelen}, \citenamefont {Marzetti},
  \citenamefont {de~Pasquale}, \citenamefont {Di~Pompeo} \emph
  {et~al.}}]{larson2013adding}%
  \BibitemOpen
  \bibfield  {author} {\bibinfo {author} {\bibfnamefont {L.~J.}\ \bibnamefont
  {Larson-Prior}}, \bibinfo {author} {\bibfnamefont {R.}~\bibnamefont
  {Oostenveld}}, \bibinfo {author} {\bibfnamefont {S.}~\bibnamefont
  {Della~Penna}}, \bibinfo {author} {\bibfnamefont {G.}~\bibnamefont
  {Michalareas}}, \bibinfo {author} {\bibfnamefont {F.}~\bibnamefont {Prior}},
  \bibinfo {author} {\bibfnamefont {A.}~\bibnamefont {Babajani-Feremi}},
  \bibinfo {author} {\bibfnamefont {J.-M.}\ \bibnamefont {Schoffelen}},
  \bibinfo {author} {\bibfnamefont {L.}~\bibnamefont {Marzetti}}, \bibinfo
  {author} {\bibfnamefont {F.}~\bibnamefont {de~Pasquale}}, \bibinfo {author}
  {\bibfnamefont {F.}~\bibnamefont {Di~Pompeo}},  \emph {et~al.},\ }\href@noop
  {} {\bibfield  {journal} {\bibinfo  {journal} {Neuroimage}\ }\textbf
  {\bibinfo {volume} {80}},\ \bibinfo {pages} {190} (\bibinfo {year}
  {2013})}\BibitemShut {NoStop}%
\bibitem [{\citenamefont {H{\"a}m{\"a}l{\"a}inen}\ \emph
  {et~al.}(1993)\citenamefont {H{\"a}m{\"a}l{\"a}inen}, \citenamefont {Hari},
  \citenamefont {Ilmoniemi}, \citenamefont {Knuutila},\ and\ \citenamefont
  {Lounasmaa}}]{hamalainen1993magnetoencephalography}%
  \BibitemOpen
  \bibfield  {author} {\bibinfo {author} {\bibfnamefont {M.}~\bibnamefont
  {H{\"a}m{\"a}l{\"a}inen}}, \bibinfo {author} {\bibfnamefont {R.}~\bibnamefont
  {Hari}}, \bibinfo {author} {\bibfnamefont {R.~J.}\ \bibnamefont {Ilmoniemi}},
  \bibinfo {author} {\bibfnamefont {J.}~\bibnamefont {Knuutila}}, \ and\
  \bibinfo {author} {\bibfnamefont {O.~V.}\ \bibnamefont {Lounasmaa}},\
  }\href@noop {} {\bibfield  {journal} {\bibinfo  {journal} {Reviews of modern
  Physics}\ }\textbf {\bibinfo {volume} {65}},\ \bibinfo {pages} {413}
  (\bibinfo {year} {1993})}\BibitemShut {NoStop}%
\bibitem [{\citenamefont {Canolty}\ \emph {et~al.}(2006)\citenamefont
  {Canolty}, \citenamefont {Edwards}, \citenamefont {Dalal}, \citenamefont
  {Soltani}, \citenamefont {Nagarajan}, \citenamefont {Kirsch}, \citenamefont
  {Berger}, \citenamefont {Barbaro},\ and\ \citenamefont
  {Knight}}]{canolty2006high}%
  \BibitemOpen
  \bibfield  {author} {\bibinfo {author} {\bibfnamefont {R.~T.}\ \bibnamefont
  {Canolty}}, \bibinfo {author} {\bibfnamefont {E.}~\bibnamefont {Edwards}},
  \bibinfo {author} {\bibfnamefont {S.~S.}\ \bibnamefont {Dalal}}, \bibinfo
  {author} {\bibfnamefont {M.}~\bibnamefont {Soltani}}, \bibinfo {author}
  {\bibfnamefont {S.~S.}\ \bibnamefont {Nagarajan}}, \bibinfo {author}
  {\bibfnamefont {H.~E.}\ \bibnamefont {Kirsch}}, \bibinfo {author}
  {\bibfnamefont {M.~S.}\ \bibnamefont {Berger}}, \bibinfo {author}
  {\bibfnamefont {N.~M.}\ \bibnamefont {Barbaro}}, \ and\ \bibinfo {author}
  {\bibfnamefont {R.~T.}\ \bibnamefont {Knight}},\ }\href@noop {} {\bibfield
  {journal} {\bibinfo  {journal} {science}\ }\textbf {\bibinfo {volume}
  {313}},\ \bibinfo {pages} {1626} (\bibinfo {year} {2006})}\BibitemShut
  {NoStop}%
\bibitem [{\citenamefont {Brookes}\ \emph {et~al.}(2011)\citenamefont
  {Brookes}, \citenamefont {Woolrich}, \citenamefont {Luckhoo}, \citenamefont
  {Price}, \citenamefont {Hale}, \citenamefont {Stephenson}, \citenamefont
  {Barnes}, \citenamefont {Smith},\ and\ \citenamefont
  {Morris}}]{brookes2011investigating}%
  \BibitemOpen
  \bibfield  {author} {\bibinfo {author} {\bibfnamefont {M.~J.}\ \bibnamefont
  {Brookes}}, \bibinfo {author} {\bibfnamefont {M.}~\bibnamefont {Woolrich}},
  \bibinfo {author} {\bibfnamefont {H.}~\bibnamefont {Luckhoo}}, \bibinfo
  {author} {\bibfnamefont {D.}~\bibnamefont {Price}}, \bibinfo {author}
  {\bibfnamefont {J.~R.}\ \bibnamefont {Hale}}, \bibinfo {author}
  {\bibfnamefont {M.~C.}\ \bibnamefont {Stephenson}}, \bibinfo {author}
  {\bibfnamefont {G.~R.}\ \bibnamefont {Barnes}}, \bibinfo {author}
  {\bibfnamefont {S.~M.}\ \bibnamefont {Smith}}, \ and\ \bibinfo {author}
  {\bibfnamefont {P.~G.}\ \bibnamefont {Morris}},\ }\href@noop {} {\bibfield
  {journal} {\bibinfo  {journal} {Proceedings of the National Academy of
  Sciences}\ }\textbf {\bibinfo {volume} {108}},\ \bibinfo {pages} {16783}
  (\bibinfo {year} {2011})}\BibitemShut {NoStop}%
\bibitem [{\citenamefont {Hillebrand}\ \emph {et~al.}(2012)\citenamefont
  {Hillebrand}, \citenamefont {Barnes}, \citenamefont {Bosboom}, \citenamefont
  {Berendse},\ and\ \citenamefont {Stam}}]{hillebrand2012frequency}%
  \BibitemOpen
  \bibfield  {author} {\bibinfo {author} {\bibfnamefont {A.}~\bibnamefont
  {Hillebrand}}, \bibinfo {author} {\bibfnamefont {G.~R.}\ \bibnamefont
  {Barnes}}, \bibinfo {author} {\bibfnamefont {J.~L.}\ \bibnamefont {Bosboom}},
  \bibinfo {author} {\bibfnamefont {H.~W.}\ \bibnamefont {Berendse}}, \ and\
  \bibinfo {author} {\bibfnamefont {C.~J.}\ \bibnamefont {Stam}},\ }\href@noop
  {} {\bibfield  {journal} {\bibinfo  {journal} {Neuroimage}\ }\textbf
  {\bibinfo {volume} {59}},\ \bibinfo {pages} {3909} (\bibinfo {year}
  {2012})}\BibitemShut {NoStop}%
\bibitem [{\citenamefont {Deuker}\ \emph {et~al.}(2009)\citenamefont {Deuker},
  \citenamefont {Bullmore}, \citenamefont {Smith}, \citenamefont {Christensen},
  \citenamefont {Nathan}, \citenamefont {Rockstroh},\ and\ \citenamefont
  {Bassett}}]{deuker2009reproducibility}%
  \BibitemOpen
  \bibfield  {author} {\bibinfo {author} {\bibfnamefont {L.}~\bibnamefont
  {Deuker}}, \bibinfo {author} {\bibfnamefont {E.~T.}\ \bibnamefont
  {Bullmore}}, \bibinfo {author} {\bibfnamefont {M.}~\bibnamefont {Smith}},
  \bibinfo {author} {\bibfnamefont {S.}~\bibnamefont {Christensen}}, \bibinfo
  {author} {\bibfnamefont {P.~J.}\ \bibnamefont {Nathan}}, \bibinfo {author}
  {\bibfnamefont {B.}~\bibnamefont {Rockstroh}}, \ and\ \bibinfo {author}
  {\bibfnamefont {D.~S.}\ \bibnamefont {Bassett}},\ }\href@noop {} {\bibfield
  {journal} {\bibinfo  {journal} {Neuroimage}\ }\textbf {\bibinfo {volume}
  {47}},\ \bibinfo {pages} {1460} (\bibinfo {year} {2009})}\BibitemShut
  {NoStop}%
\bibitem [{\citenamefont {Telesford}, \citenamefont {Burdette},\ and\
  \citenamefont {Laurienti}(2013)}]{telesford2013exploration}%
  \BibitemOpen
  \bibfield  {author} {\bibinfo {author} {\bibfnamefont {Q.~K.}\ \bibnamefont
  {Telesford}}, \bibinfo {author} {\bibfnamefont {J.~H.}\ \bibnamefont
  {Burdette}}, \ and\ \bibinfo {author} {\bibfnamefont {P.~J.}\ \bibnamefont
  {Laurienti}},\ }\href@noop {} {\bibfield  {journal} {\bibinfo  {journal}
  {Frontiers in neuroscience}\ }\textbf {\bibinfo {volume} {7}} (\bibinfo
  {year} {2013})}\BibitemShut {NoStop}%
\bibitem [{\citenamefont {Rocha}, \citenamefont {Masuda},\ and\ \citenamefont
  {Holme}(2017)}]{rocha2017sampling}%
  \BibitemOpen
  \bibfield  {author} {\bibinfo {author} {\bibfnamefont {L.~E.~C.}\
  \bibnamefont {Rocha}}, \bibinfo {author} {\bibfnamefont {N.}~\bibnamefont
  {Masuda}}, \ and\ \bibinfo {author} {\bibfnamefont {P.}~\bibnamefont
  {Holme}},\ }\href@noop {} {\bibfield  {journal} {\bibinfo  {journal}
  {Physical Review E}\ }\textbf {\bibinfo {volume} {96}},\ \bibinfo {pages}
  {052302} (\bibinfo {year} {2017})}\BibitemShut {NoStop}%
\bibitem [{\citenamefont {Jeh}\ and\ \citenamefont
  {Widom}(2002)}]{jeh2002simrank}%
  \BibitemOpen
  \bibfield  {author} {\bibinfo {author} {\bibfnamefont {G.}~\bibnamefont
  {Jeh}}\ and\ \bibinfo {author} {\bibfnamefont {J.}~\bibnamefont {Widom}},\
  }in\ \href@noop {} {\emph {\bibinfo {booktitle} {Proceedings of the eighth
  ACM SIGKDD international conference on Knowledge discovery and data
  mining}}}\ (\bibinfo {organization} {ACM},\ \bibinfo {year} {2002})\ pp.\
  \bibinfo {pages} {538--543}\BibitemShut {NoStop}%
\end{thebibliography}%

\end{document}